\documentclass{article}[12pt]

\usepackage{graphicx}

\begin{document}

\noindent {\em Astrophysics, Vol. 58, No. 3, September, 2015\\
Original article submitted May 7, 2015. Translated from Astrofizika, Vol. 58, No. 3, pp. 331-339 (August
2015).\\
0571-7256/15/5803-0309 @2015 Springer Science+Business Media New York}

\bigskip

\begin{center}

{\bf \LARGE NEW RADIAL VELOCITIES FOR DWARF GALAXIES IN THE LOCAL VOLUME}

\bigskip

{\bf I. D. Karachentsev, M. E. Sharina, D. I. Makarov, Yu. A. Perepelitsyna, and E. S. Safonova}

\bigskip

{\bf Special Astrophysical Observatory, Russian Academy of Sciences; e-mail: ikar@sao.ru}

\end{center}

\bigskip

{\em Radial velocities measured with the 6-meter telescope are given for 5 faint dwarf galaxies. All of these
galaxies are confirmed as very nearby objects. Two of them, KK135 (dIr) and UGC 1703 (dSph/dTr), are
local isolated dwarfs, and the three others, UGCA 127sat (dIr), NGC 2683dw1 (dIr), and NGC891dwA
(dTr), belong to companions of nearby massive spirals.}

\bigskip

{\bf 1. Introduction}

\bigskip

In the astronomical literature there are a number of catalogs of galaxies which are organized in terms of a
limiting flux of the objects in the optical and infrared ranges, or in terms of the flux in the 21 cm radio line [1--4].
Kraan-Korteweg and Tammann [5] provided a justification for a catalog based on distance, rather than flux (apparent
magnitude). The need for a representative sample of nearby galaxies within a fixed volume is dictated as well by
the need to compare various ``mock'' catalogs produced by numerical modelling of large-scale structures of the
universe [6,7] with observational data. The catalog of Ref. 5 enumerated a total of 179 galaxies with expected
distances within a range of 10 Mpc. The obvious difficulty of creating such a catalog arose from the absence at that
time of reliable distance estimates, even for the closest and brightest galaxies. The situation has changed 
fundamentally over the last two decades, when bulk distance measurements began to be made with the Hubble telescope. The
progress of these measurements, which are based on the luminosity of stars at the top of the red giant branch, is
illustrated in the Catalog of Neighboring Galaxies, CNG [8], and in the Extragalactic Distance Database, EDD [9]. The revised
and supplemented UNGC catalog [10] contains 869 galaxies, or almost five times the number in the original sample
[5]. The UNGC catalog includes galaxies with radial velocities $V_{LG} < 600$ km/s relative to the centroid of the Local
group, or galaxies with individual estimated distances $D < 11$ Mpc.

At the time the UNGC was published (2013), half of the galaxies in the catalog of distances were measured
with an accuracy of better than 10\%. At the same time, data on radial velocity are lacking for 108 of the galaxies
(12\% of the sample). Most of these are distinguished by a low surface brightness, and do not contain HII regions
or a significant amount of neutral hydrogen. Objects of this sort with old star populations usually occure near highly
luminous galaxies subject to gas sweeping from the shallow potential well of a dwarf as it interacts with a massive
neighbor. In recent years, there has been a rapid increase in the number of galaxies with low surface brightnesses
in the Local volume ($D < 11$ Mpc) because of targeted searches for dwarf companions around massive nearby galaxies,
which can be photographed with small-diameter telescopes but with very long exposure times [11--13].

Among the dwarf galaxies without radial velocities there are a few that are not associated with bright galaxies.
The UNGC catalog contains a total of 17 such objects. Two of these, KK~258 and KKs3, turn out to be very nearby,
isolated systems at distances of $\sim$2 Mpc [14, 15]. Measurement of the radial velocities of this small category of
galaxies is an important step toward understanding their nature. We emphasize that the origin of isolated dwarf 
spheroidal(dSph) galaxies has not yet been explained convincingly in modern cosmological models.

\bigskip

{\bf 2. Observations and data processing}

\bigskip

At the Special Astrophysical Observatory of the Russian Academy of Sciences, spectral observations have been
made on the 6-meter telescope of galaxies within the Local volume for which the radial velocities have not been
estimated before. The SCORPIO universal focal reducer was used in the ``long slit'' mode [16] with a spectral
resolution of $\sim10$\AA/pixel. Table 1 is a list of the observed galaxies, each with an observation date, exposure time,
average seeing, and the diffraction grating that was employed.

The spectra recorded with the CCD were reduced in the standard MIDAS system using the LONG package.
The radial velocity of an object was determined in two ways: by measuring individual emission or absorption lines
and by cross correlation with a reference spectrum.

The results of our observations are listed in Table 2. The first and second columns give the name of the galaxy
and its coordinates at the epoch J2000.0, and the third through seventh columns give the integrated B magnitude,
morphological type, absolute magnitude, and the heliocentric radial velocity and the corresponding measurement
error.

\begin{table}[hbt]
\begin{center}
\caption{\label{tab:log} Observation Log}
\begin{tabular}{|l|c|c|c|l|}
\hline\hline
Galaxy & Date        & Exposure & Seeing   &  Grating  \\
         &  & (s)    &  $^{\prime\prime}$ &  \\
\hline 
KK135        & 25.02.2015 & 2 x 1200 & 1.7 &  VPHG550G  \\
UGCA~127~sat& 28.10.2014 & 300, 900  & 1.3 &  VPHG1200B \\
UGC1703    & 26.10.2014 & 4 x 1200  & 1.2 &   VPHG1200B \\
UGC1703       & 28.10.2014 & 8 x 1200  & 1.1 &   VPHG1200B \\
NGC~2683~dw1 & 29.10.2014 & 2 x 900  & 0.9 &   VPHG1200B   \\
$[$TT2009$]$25   & 19.02.2015 & 3 x 1200 & 3.0 &   VPHG1200G  \\
NGC672 dwD  & 23.01.2015  & 2 x 600  & 1.0 &   VPHG1200R \\
\hline\hline
\end{tabular}
\end{center}
\end{table}

\begin{table}[hbt]
\begin{center}
  \caption{Parameters of the Observed Galaxies}
  \begin{tabular}{|l|c|c|c|r|r|r|} \hline
   Galaxy     &    RA (J2000.0) Dec &   $B $ & Type& $M_B$&  $V_h$  & $\sigma_V$   \\
                  &                    & mag&    & & km/s & km/s\\
   \hline
   KK~135    &   121934.7+580234 &  18.1  & dIr & -10.2& 215 &   40 \\
 UGCA~127~sat  & 062054.8--083901   &16.9 & dIr & -16.3&  708   & 26  \\
 UGC~1703     & 021255.8+324851   &17.0 &dSph & -11.5   &40   & 20   \\
 NGC~2683~dw1  & 085326.8+331820   &19.5 & dIr & -12.0  &380   & 25   \\
 $[$ТТ2009$]$25  & 022112.4+422150   &17.9 &dTr &-12.3  &692   & 58  \\
 NGC~672~dwD   & 014738.4+272620   &18.7& S & -19.6 &29860   &110     \\
 \hline
 \end{tabular}
 \end{center}
 \end{table}
 
 \bigskip
 
 {\bf 3. Discussion of results}
 
 \bigskip
 
Here we briefly note some features of the observed galaxies.

{\bf 3.1. KK 135=PGC 166130=SDSS J121934.68+580234.4.} This irregular dwarf galaxy with an angular size
of $0.68^{\prime} \times 0.35^{\prime}$ was reported in Ref. 17, but not included in the UNGC catalog because the 
radial velocity had not
been measured. In the ultraviolet sky survey GALEX [18] it has apparent magnitudes of $m(FUV ) = 18.70\pm0.12$ and
$m(NUV ) = 18.60\pm0.04$. An image of this galaxy from the Sloan Digital Sky Survey [19] is shown in Fig. 1. A
reproduction of the spectrum of KK~135 obtained by us in the 4000-7000\AA \,range is shown in Fig. 2. Judging from
the measured radial velocity $V_h = 215\pm40$ km/s, this galaxy certainly belongs to the Local volume sample. A
compact HII region lies on the southern side of KK~135. KK~135 is an extremely isolated galaxy. Another irregular
galaxy DDO~123 lies in its neighborhood at an angular separation of $0.9^{\circ}$, but that galaxy has a radial velocity
$V_h = 722$ km/s and is at a distance of $D = 10.5$ Mpc [10], so there is no physical coupling of these systems.

\begin{figure}[hbt]
\begin{center}
\includegraphics[scale=0.3]{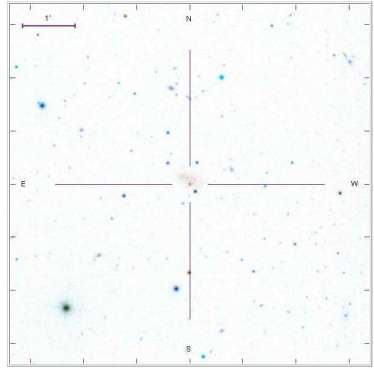}
\caption{A reproduction of the dwarf galaxy KK~135 from the SDSS survey.}
\end{center}
\end{figure}

\begin{figure}[hbt]
\begin{center}
\includegraphics[scale=0.3,angle=-90]{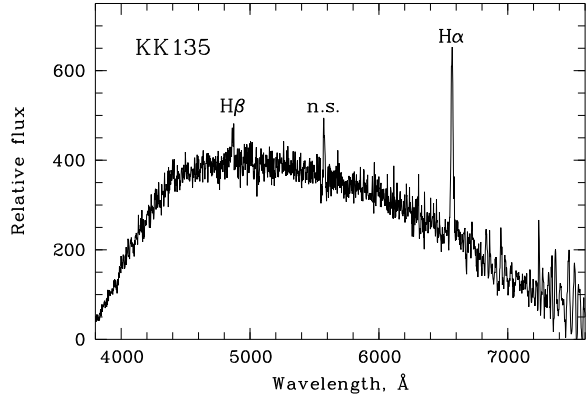}
\caption{Spectrum of the dwarf galaxy KK 135.}
\end{center}
\end{figure}

\begin{figure}
\begin{center}
\includegraphics[scale=0.3,angle=-90]{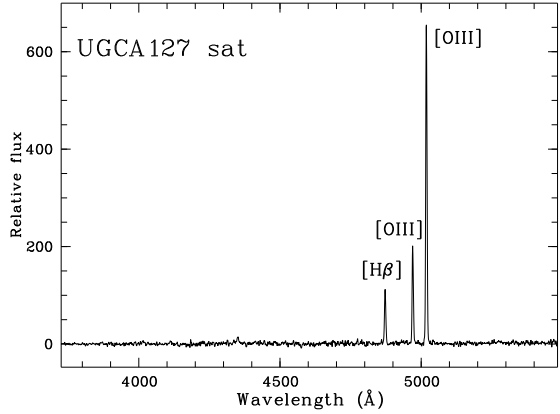}
\caption{Spectrum of the dwarf galaxy UGCA 127 sat.}
\end{center}
\end{figure}

\bigskip

{\bf 3.2. UGCA 127 sat=LV J0620-0839.} This dIr galaxy lies in a zone of strong galactic absorption ($A_b = 3.53^m$)
10$^{\prime}$ to the south of the spiral galaxy UGCA~127. Bright emission lines can be seen in the spectrum of this presumed
dwarf companion (Fig. 3); from these we measured a radial velocity for the companion of $V_h = 708\pm26$ km/s, which
is close to the velocity of the neighboring spiral galaxy, $V_h = 732$ km/s, and confirms the physical coupling of the
two. To the east of UGCA 127 sat, there is an emission node [20] with a radial velocity 80 km/s higher than that
of the central part of the galaxy.

\begin{figure}[hbt]
\begin{center}
\includegraphics[scale=0.3,angle=-90]{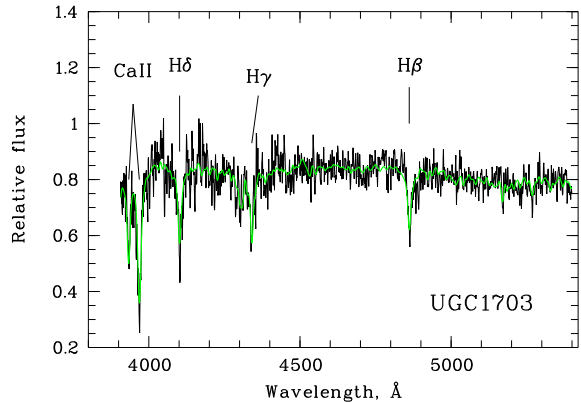}
\caption{Spectrum of the dwarf galaxy UGC 1703.}
\end{center}
\end{figure}

\begin{figure}[hbt]
\begin{center}
\includegraphics[scale=0.3,angle=-90]{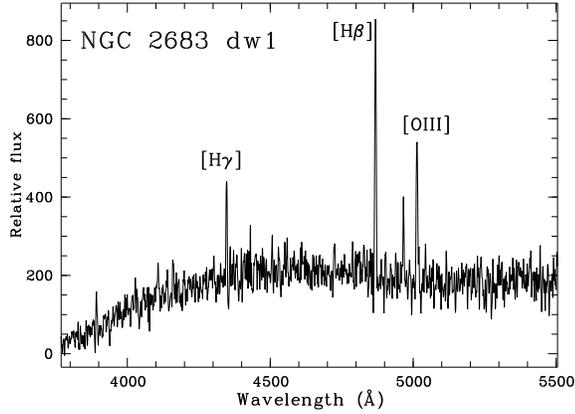}
\caption{Spectrum of the dwarf galaxy NGC 2683 dw1.}
\end{center}
\end{figure}

\bigskip

{\bf 3.3. UGC 1703 = KKH 9 = PGC 08484.} This is a dwarf galaxy of spheroidal (dSph) or transition (dTr) type.
The distance $D = 4.19$ Mpc to it has been estimated from the fluctuations in its surface brightness [21]. In an 
$H_{\alpha}$ line trace [22] it shows no signs of emission. Absorption $H_{\beta}$,
$H_{\gamma}$ , and $H_{\delta}$ Balmer lines and the CaII (H, K) doublet
with an average radial velocity of +40 km/s can be seen in its spectrum (Fig. 4).

For this distance the absolute magnitude of UGC~1703 is $M_B = -11.5^m$. This galaxy belongs to a rare class
of isolated dwarf systems with an old star population. Its nearest massive neighbor is the spiral galaxy Maffei 2.

\bigskip

{\bf 3.4. NGC 2683 dw1.} This faint, irregular galaxy was discovered in an image of the neighborhood of NGC
2683 obtained using a small telescope by M.~Elvov with an exposure of 15 hours [23]. It was detected in the GALEX
survey [18] as a faint UV source. Its spectrum (Fig. 5) contains [OIII], $H_{\beta}$, and $H_{\gamma}$
 emission lines, which imply a radial velocity for the dwarf object of $V_h = 380\pm25$ km/s. 
 This confirms that it is physically coupled to the spiral galaxy
NGC 2683, which has a velocity Vh = 411 km/s and a distance of 9.36 Mpc [24]. Therefore, NGC 2683 dw1 is a
new representative of the galaxies of the Local volume.

\begin{figure}
\begin{center}
\includegraphics[scale=0.3,angle=-90]{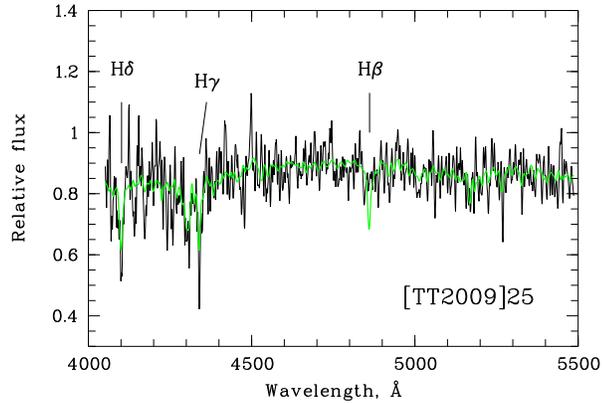}
\caption{Spectrum of the dwarf galaxy [TT2009]25, companion of NGC 891.}
\end{center}
\end{figure}

\bigskip

{\bf 3.5. [TT2009]25 = NGC 891 dwA.} This is one of two dwarf galaxies with low surface brightnesses discovered
[23] in the neighborhood of the spiral galaxy NGC~891. Even earlier it was noted as a probable companion of NGC~891
 in a survey with the MegaCam on the 3.6-meter CFHT telescope [25]. Judging from the smooth structure and
the presence of a faint UV flux, [TT2009]25 can be classified as a dwarf system of type dTr, intermediate between
dSph and dIr. Only absorption lines show up in its spectrum; this indicate a heliocentric velocity for this galaxy
of $692\pm58$ km/s, which is consistent with its status as a companion of the spiral NGC~891, for which
$V_h = 526$ km/s.

\bigskip

{\bf 3.6. NGC 672 dwD.} During a search for new companions surrounding nearby massive galaxies [23], a faint
bluish compact object with a visible magnitude of $B = 18.7^m$ was noticed. We obtained its spectrum and found that
it is actually a distant spiral galaxy with radial velocity $V_h = 29860\pm110$ km/s.

\bigskip

{\bf 4. Concluding comment}

\bigskip

Searches for new ultrafaint companions around nearby massive galaxies by amateur astronomers with small
telescopes are extremely important for the cosmology of the Local universe. This program augments the sample of
``test particles'' in the Local volume, so that the distribution of the gravitational potential can be tested on
small scales with an unprecedented high density of data. Bulk searches for candidate members of the Local volume must
obviously be accompanied by measurements of their radial velocities on large telescopes. The combined efforts of
amateur astronomical photographers and professional astronomers promise great advances in the study of the Local
velocity field of galaxies.
This work was supported by a grant from the Russian Science Foundation (project No. 14--12--00965).

\bigskip

{\bf REFERENCES}

\bigskip

1. A. Sandage and G. A. Tammann, Revised Shapley-Ames Catalog of Bright Galaxies, Carnegie Institution,
Washington, DC, p. 635 (1981).

2. F. Zwicky, et al., Catalog of Galaxies and Clusters of Galaxies, Caltech, Pasadena (1961).

3. T. H. Jarrett, T. Chester, R. Cutri, S. E. Schneider, and J. P. Huchra, Astron. J. 125, 525 (2003).

4. B. S. Koribalski, L. Staveley-Smith, V. A. Kilborn, et al., Astron. J. 128, 16 (2004).

5. R. C. Kraan-Korteweg and G. A. Tammann, Astron. Nachr. 300, 181 (1979).

6. M. Boylan-Kolchin, V. Springel, S. D. M. White, et al., Mon. Not. Roy. Astron. Soc. 398, 1150 (2009).

7. S. E. Nuza, F. S. Kitaura, S. Hess, et al., Mon. Not. Roy. Astron. Soc. 445, 988 (2014).

8. I. D. Karachentsev, V. E. Karachentseva, W. K. Huchtmeier, and D. I. Makarov, Astron. J. 127, 2031 (=CNG),
(2004).

9. B. A. Jacobs, L. Rizzi, R. B. Tully, et al., Astron. J. 138, 332 (2009).

10. I. D. Karachentsev, D. I. Makarov, and E. I. Kaisina, Astron. J. 145, 101 (=UNGC), (2013).

11. P. G. van Dokkum, R. Abraham, and A. Merritt, Astron. J. 782, 24 (2014).

12. I. D. Karachentsev, D. Bautzmann, F. Neyer, et al., arXiv:1401. 2719 (2014).

13. A. Merritt, P. van Dokkum, and R. Abraham, Astrophys. J. 787, 37L (2014).

14. I. D. Karachentsev, L. N. Makarova, D. I. Makarov, et al., Mon. Not. Roy. Astron. Soc. 447L, 85 (2015).

15. I. D. Karachentsev, L. N. Makarova, R. B. Tully, et al., Mon. Not. Roy. Astron. Soc. 443, 1281 (2014).

16. V. L. Afanasjev, E. B. Gazhur, S. R. Zhelenkov, and A. V. Moiseev, Astrophys. Bulletin 58, 90 (2005).

17. V. E. Karachentseva and I. D. Karachentsev, Astron. Astrophys. Suppl. Ser. 127, 409 (1998).

18. A. Gil de Paz, B. F. Madore, S. Boissier, et al., Astrophys. J. 627L, 29 (2005).

19. K. N. Abazajian, J. K. Adelman-McCarthy, M. A. Agueros, et al., Astrophys. J. Suppl. 182, 54 (2009).

20. S. S. Kaisin, I. D. Karachentsev, and S. Ravindranath, Mon. Not. Roy. Astron. Soc. 425, 2083 (2012).

21. R. Rekola, H. Jerjen, and C. Flynn, Astron. Astrophys. 437, 823 (2005).

22. I. D. Karachentsev and S. S. Kaisin, Astron. J. 140, 1241 (2010).

23. I. D. Karachentsev, D. Bautzmann, M. Blauensteiner, et al., Astrophys. Bulletin (2015) (accepted).

24. I. D. Karachentsev, R. B. Tully, L. N. Makarova, et al., Astrophys. J. 805, 144 (2015).

25. N. Trentham and R. B. Tully, Mon. Not. Roy. Astron. Soc. 398, 722 (2009).

\end{document}